\documentclass[final]{colt2026}

\usepackage{times}
\usepackage{mathtools}

\newtheorem{problem}[theorem]{Open Problem}

\title[Is Interaction Necessary for Order-Optimal 1-bit Mean Estimation?]{Open Problem: Is Interaction Necessary \\ for Order-Optimal 1-bit Mean Estimation?}

\coltauthor{%
 \Name{Ivan Lau} \Email{ivan.lau@u.nus.edu}\\
 \addr National University of Singapore
 \AND
 \Name{Jonathan Scarlett} \Email{scarlett@comp.nus.edu.sg}\\
 \addr National University of Singapore \\
}

\begin{document}

\maketitle

\begin{abstract}
We ask whether interaction is necessary for order-optimal 1-bit mean
estimation over nonparametric finite-moment classes.  Adaptive threshold-query protocols achieve the order-optimal 1-bit minimax rate, and the same rate is attainable with general 1-bit queries using only one adaptive transition (i.e., two stages of querying).  In the non-adaptive setting, threshold and interval queries are known to be highly suboptimal, but the case of arbitrary
non-adaptive quantizers remains unresolved.  Can such quantizers match
the adaptive rate, yielding an optimal one-shot protocol?  Or is the known two-stage estimator stage-optimal, with a single adaptive transition being necessary and sufficient?
\end{abstract}


\section{Motivation}

Mean estimation is one of the most fundamental primitives in statistics, machine learning, and theoretical computer science. In many modern data-acquisition systems, samples are not observed directly by the learner. Instead, they are held by agents, such as low-power sensors, edge devices, or clients in a federated system, that may transmit only a few bits per observation.  A fundamental limiting case is 1-bit communication: each sample is compressed to a single binary message before being transmitted to the learner. 

At first glance, one bit per sample may appear too restrictive for
nonparametric inference.  However, recent work shows that this intuition is
misleading.  With interaction/adaptivity (i.e., 1-bit queries that may be designed based on the previous bits observed), 1-bit mean estimation remains nearly as sample-efficient as \emph{unquantized} mean estimation: adaptive 1-bit estimators match the unquantized minimax rate up to at most one (unavoidable) logarithmic factor. In fact, simple adaptive threshold queries already suffice \citep{lau2026order}, and the resulting rates are order-optimal in the 1-bit model.  However, the 1-bit constraint itself is not the
whole story, and a more refined question is: \emph{Is interaction/adaptivity truly necessary to achieve this order-optimal rate?}

We study this question in a nonparametric setting.  The learner receives
one bit per sample and must estimate the mean of an unknown distribution
whose mean lies in a known bounded interval and whose $k$-th central
moment is bounded.  The quantizers may either be chosen \emph{adaptively}, using
previously received bits, or fixed \emph{non-adaptively} before any messages
are observed.  The fully non-adaptive model is operationally appealing: devices can be pre-programmed and communicate without feedback or multiple
adaptive rounds.

The current landscape is sharp but incomplete.  Fully non-adaptive
threshold and interval queries are known to be highly suboptimal~\citep{lau2025sequential}, i.e., these query families cannot match the adaptive rate.  However, this
does not rule out non-adaptive protocols using general 1-bit quantizers, which are potentially much more expressive.  If just one adaptive
transition is allowed, then the order-optimal
rate is known to be achievable using general 1-bit queries~\citep{lau2026order}.  The unresolved case is therefore precisely the
zero-adaptivity, general-query regime:
\begin{quote}
\emph{Can fully non-adaptive arbitrary 1-bit quantizers match the
adaptive 1-bit minimax rate for nonparametric mean estimation, or is
interaction necessary?}
\end{quote}
Either answer would be significant. A positive answer would give an order-optimal non-adaptive protocol for
nonparametric 1-bit mean estimation.  A negative answer would establish a genuine interaction/adaptivity gap, and certify that the known two-stage estimator has the fewest stages possible: one adaptive transition suffices, but zero adaptive transitions do not.

\section{Problem Setup and Open Problem}
\label{sec:setup}

\textbf{Distributional assumption.}
We consider one-dimensional mean estimation over the nonparametric distribution family
\[
    \mathcal D(k,\lambda,\sigma)
    =
    \left\{
        D:\ \mu(D) \coloneqq \mathbb E_{X\sim D}[X]\in[-\lambda,\lambda],
        \quad
        \mathbb E_{X\sim D} |X-\mu(D)|^k\le \sigma^k
    \right\},
\]
where $k>1$ and $\lambda\ge \sigma>0$ are known to the learner.  Apart from the tail/moment condition, the class is unrestricted: the distribution
may be discrete, asymmetric, or supported on an unbounded
set.  The parameter $\lambda$ is a search radius for the mean, while
$\sigma$ represents the intrinsic scale (e.g., the standard deviation when $k=2$).

\paragraph{1-bit protocols.}
A 1-bit communication protocol observes independent samples $X_1,\ldots,X_n \sim D$ only through binary messages $Y_t=\mathbf{1}\{X_t\in A_t\}$, where
$A_t \subseteq\mathbb R$ is measurable. In an adaptive protocol,  $A_t$ may be chosen based on the entire past transcript $(A_1, Y_1,\ldots, A_{t-1}, Y_{t-1})$ as well as any internal randomness.  In a fully non-adaptive protocol, all sets $A_1,\ldots,A_n$ are fixed before any messages are observed, possibly using public or private randomness.
Threshold and interval queries correspond to $A_t$ being a half-line or an interval, respectively. General quantizers/queries can be highly nonlocal: for example, 
% after partitioning $[-\lambda,\lambda]$ into bins and assigning binary codewords to bins, a query may reveal one codeword coordinate, so its preimage is a union of many separated bins.
coding-based queries may have preimages that are unions of many intervals.
We say that a protocol is $(\epsilon,\delta)$-accurate over
$\mathcal D(k,\lambda,\sigma)$ if its output $\widehat\mu$ satisfies
\[
    \sup_{D \in\mathcal D(k,\lambda,\sigma)}
    \mathbb P\left\{
        |\widehat\mu-\mu(D)|> \epsilon
    \right\}
    \le \delta ,
\]
where the probability is over the samples and any internal randomness of
the protocol.

\paragraph{Order-optimal rate.}
For $\epsilon$ below a sufficiently small constant multiple of $\sigma$
and $\delta\in(0,1/2)$, the adaptive 1-bit minimax sample complexity is
known, up to constants depending only on $k$, to be
\begin{equation}\label{eq:rate}
    r_k(\lambda, \sigma,\epsilon,\delta) \coloneqq
    \underbrace{\log\frac{\lambda}{\sigma}}_{\text{localization}}
    +
    \underbrace{
    \begin{cases}
    \displaystyle
    \frac{\sigma^2}{\epsilon^2} \cdot \log\frac1\delta,
    & k>2, \\[0.8em]
    \displaystyle
    \frac{\sigma^2}{\epsilon^2} \cdot
    \log\frac{\sigma}{\epsilon} \cdot
    \log\frac1\delta,
    & k=2, \\[0.8em]
    \displaystyle
    \left(\frac{\sigma}{\epsilon}\right)^{k/(k-1)} \cdot
    \log\frac1\delta,
    & 1<k<2 .
    \end{cases}
    }_{\text{refinement}}
\end{equation}
This rate is attainable by adaptive 1-bit protocols and is minimax
optimal among 1-bit protocols~\citep{lau2026order}.
The additive term $\log(\lambda/\sigma)$ corresponds to the cost of \textit{localizing} the mean from $[-\lambda,\lambda]$ to an interval of length $O(\sigma)$. The remaining term is the \textit{refinement} cost once such an interval has been identified.  For $k\neq 2$, this matches the usual unquantized minimax rate up to the localization term.  For $k=2$, the extra $\log(\sigma/\epsilon)$ factor is an unavoidable 1-bit penalty.  
The open problem is whether fully non-adaptive arbitrary 1-bit quantizers
can achieve the same rate.
% We now state our open problem formally, which asks whether fully non-adaptive arbitrary 1-bit quantizers can achieve the same rate.
\begin{problem}[Order-optimal non-adaptive 1-bit mean estimation]
\label{prob:main}
Fix $k>1$. Do there exist constants $c_k,C_k>0$ such that, for all
$\lambda\ge\sigma>0$, all $0<\epsilon\le c_k\sigma$, and all
$\delta\in(0,1/2)$, there is a fully non-adaptive 1-bit protocol that
is $(\epsilon,\delta)$-accurate over $\mathcal D(k,\lambda,\sigma)$ using
at most $n \le C_k\, r_k(\lambda,\sigma,\epsilon,\delta)$
samples?
\end{problem}
\section{Prior Work and Known Results}\label{sec:prior-work}

\paragraph{Classical mean estimation.}
Without communication constraints, high-probability mean estimation under
moment assumptions is well understood. For $k \ge 2$, the optimal sample complexity is of order $(\sigma^2/\epsilon^2) \cdot \log(1/\delta)$; for $1 < k  < 2$, it is of order $(\sigma/\epsilon)^{\frac{k}{k-1}} \cdot \log(1/\delta)$ --
see~\citep{devroye2016sub,lee2022optimal, cherapanamjeri2022optimal,minsker2023efficient,dang2023optimality}
and the references therein. These unquantized rates are the baseline for asking how much additional sample complexity is caused by 1-bit communication and by the absence of interaction.

\paragraph{Communication constraints and interaction.}
Communication-constrained estimation has been studied extensively, with the goal of characterizing the minimax rate when each agent can transmit
only a limited number of bits~\citep{ZhangDJW13, Shamir14, BravermanGMNW16, HanOW18, duchi2019lower, BarnesHO20}.  A recurring question is whether interaction changes the statistical complexity. The answer is problem-dependent: interaction can improve sample complexity in tasks such as structured high-dimensional estimation, large-margin learning, and locally private hypothesis
selection~\citep{dagan2020interaction, gopi2020locally, acharya2022role,acharya2023unified,pour2024sample}, whereas in some distributed testing problems it gives no improvement beyond constants \citep{kazemi2025sample}. These works
motivate Problem~\ref{prob:main}, but they do not resolve it: the present setting is scalar rather than high-dimensional, nonparametric rather than a fixed parametric model, and constrained to exactly one bit per sample.

\paragraph{Parametric 1-bit mean estimation.}
Order-optimal non-adaptive 1-bit protocols are known for parametric families, including Gaussian, symmetric log-concave,
and scale-location models~\citep{kipnis2022mean, cai2024distributed, kumar2025one}. These results show that non-adaptive 1-bit estimation can be order-optimal when the distribution has exploitable structure.  For example, in Gaussian mean estimation, Gray-code-type quantizers can encode location information across multiple spatial scales \citep{cai2024distributed}.  However, these protocols rely on model-specific information such as likelihood shape, symmetry, parametric CDFs, or scale-location structure, and do not apply to the nonparametric class $\mathcal D(k,\lambda,\sigma)$.

\paragraph{Nonparametric 1-bit mean estimation.}
For $\mathcal D(k,\lambda,\sigma)$, the adaptive case is settled.
\citet[Theorem 5]{lau2026order} give an adaptive threshold-query estimator with order-optimal sample complexity~\eqref{eq:rate}. Although this estimator uses only (randomized) threshold queries, its noisy binary search based localization step is sequential, requiring $O(\log(\lambda/\sigma))$ adaptive localization steps.   
Allowing general 1-bit queries greatly reduces this adaptivity requirement:
\citet[Section~4.3]{lau2026order} give an order-optimal two-stage protocol whose first stage non-adaptively localizes the mean to an
$O(\sigma)$ interval using coding-based queries, and whose second stage,
chosen \emph{after} this interval is decoded, refines the estimate within
it.
% Thus, at least for the framework in \citep{lau2026order}, the use of general queries reduces the interaction requirement from $O\left(\log (\lambda/\sigma)\right)$ to one adaptive transition.
Thus the two-stage estimator achieves
order-optimality by separating localization from location-dependent refinement.
Problem~\ref{prob:main} asks whether this
single adaptive transition is truly necessary.

The fully non-adaptive case is less understood.  For $k=2$, stochastic quantization, equivalently randomized threshold querying, gives some non-adaptive upper bounds.  Along the lines of \citet{abdalla2026robust}, truncating to a range of size $B = \Theta(\max\{\lambda, \sigma^2/\epsilon  \})$ and then stochastically quantizing gives sample complexity of order $O((B^2/\epsilon^2) \cdot \log(1/\delta))$, which is at least quadratic in~$\lambda$. On the lower bound side, \citet{lau2025sequential} show that non-adaptive threshold and interval queries are suboptimal: their sample complexity must scale linearly with $\lambda/\sigma$, whereas adaptive threshold queries incur only logarithmic dependence. This adaptivity gap, however, does not rule out arbitrary measurable 1-bit queries.
For example, consider i.i.d. randomized Fourier-feature queries
$Z_t=2\mathbf{1}\{\cos(S_tX_t+\Theta_t)\ge U_t\}-1$, where $S_t\sim{\rm Unif}[-a,a]$, $\Theta_t\sim{\rm Unif}[0,2\pi]$, and $U_t\sim{\rm Unif}[-1,1]$. 
Let $G_\epsilon$ be an $O(\epsilon)$-spaced grid of $[-\lambda,\lambda]$ and decode by $\widehat{\mu}\in\arg\min_{\nu\in G_\epsilon}\sum_{t=1}^n (Z_t-\cos(S_t\nu+\Theta_t))^2$.
The main population gap at distance $\epsilon$ is $\Omega(a^2\epsilon^2)$, while a direct second-order bias bound gives $|\mathbb{E}[Z_t\mid S_t,\Theta_t]-\cos(S_t\mu+\Theta_t)|=O(S_t^2\sigma^2)$. Thus this analysis requires $a\lesssim \epsilon/\sigma^2$, and Hoeffding's inequality plus a union bound over $|G_\epsilon|=\Theta(\lambda/\epsilon)$ gives $O((\sigma/\epsilon)^8(\log(\lambda/\epsilon)+\log(1/\delta)))$ samples.
% Any grid point at distance at least $\epsilon$ from $\mu$ has population excess squared loss $\Omega(a^2\epsilon^2)$ relative to the grid point nearest to $\mu$. However, since $\mathbb{E}[Z_t\mid S_t,\Theta_t] = \mathbb{E}[\cos(S_tX_t+\Theta_t)]$  differs from $\cos(S_t\mu+\Theta_t)$ by $O(S_t^2\sigma^2)$, bias control requires $a = O(\epsilon/\sigma^2)$. Hoeffding's inequality and a union bound over  $|G_\epsilon| = \Theta(\lambda/\epsilon)$ yields a sample complexity of $O((\sigma/\epsilon)^8(\log(\lambda/\epsilon)+\log(1/\delta)))$. 
Hence the scheme has the desired logarithmic dependence on~$\lambda$, but a highly
suboptimal refinement rate compared to~\eqref{eq:rate}. Overall, known non-adaptive 1-bit estimators use i.i.d. queries that do not provide refinement queries tailored to the localized interval, which seems to be the cause of poor scaling in either $\lambda$ or $\sigma/\epsilon$.

% The main population gap at distance
% $\epsilon$ is $\Omega(a^2\epsilon^2)$, while a direct second-order bias
% bound gives
% $|\mathbb E[Z_t\mid S_t,\Theta_t]-\cos(S_t\mu+\Theta_t)|
% =O(S_t^2\sigma^2)$, requiring $a\lesssim\epsilon/\sigma^2$.

\section{Discussion and Technical Barriers}
\label{sec:barriers}

The known results leave a delicate gap.  In the two-stage estimator of~\citet{lau2026order}, the only adaptive transition occurs when the refinement-query block is chosen after a coarse interval is decoded from the localization bits. 
A natural route to a positive answer to Problem~\ref{prob:main} would therefore be a universal refinement scheme that is valid across all possible localized intervals.  A fully non-adaptive protocol must choose all refinement queries in advance; only the final decoder
may use the decoded interval to ``reinterpret'' the bits.  For very light-tailed classes, one might hope to use periodic quantizers so that, after the coarse location is decoded, the same bits can be reinterpreted as local refinement information.  For the considerably larger class $\mathcal D(k,\lambda,\sigma)$, however, the main difficulty is tail control.  Under only moment assumptions, far-tail samples can alias into central regions of a fixed global query and corrupt small-scale estimates.  The two-stage estimator avoids this by centering tail-sensitive refinement queries around the localized interval.  A fully non-adaptive estimator would need comparable local variance control
without knowing the interval in advance.

Conversely, a negative answer would require a lower bound ruling out such universal refinement schemes.  Existing lower bounds for interval queries exploit locality: a non-adaptive protocol must spread
informative tests across the $\Theta(\lambda/\sigma)$ possible  mean locations.  Arbitrary measurable quantizers do not obey this locality constraint; a single query can encode many locations at
once, as in the coding-based localization stage of
\citet{lau2026order}.  Hence standard packing arguments are insufficient: one would need to prove that simultaneous non-adaptive localization, refinement, and tail control necessarily incur an additional order-wise sample cost.

\newpage
\bibliography{bibliography}

\end{document}